\documentclass[journal,12pt,onecolumn,draftclsnofoot,]{IEEEtran}

\usepackage{graphicx} 
\graphicspath{./Pics/}
\DeclareGraphicsExtensions{.eps}
\usepackage{epstopdf}

\usepackage{amsmath}
\usepackage{amssymb}
\usepackage{amsfonts}
\DeclareMathOperator{\tr}{tr}
\DeclareMathOperator{\diag}{diag}

\usepackage{algorithm}
\usepackage{algorithmic}
\usepackage{color}
\usepackage{subfigure}

\usepackage[compress]{cite}
\makeatletter
\renewcommand{\citepunct}{,\penalty\@m\hskip.13emplus.1emminus.1em}
\renewcommand{\citedash}{\hbox{--}\penalty\@m}
\makeatother

\ifCLASSINFOpdf

\else
\fi

\begin{document}
\title{Hybrid Beamforming Structure for Massive MIMO System: Full-connection v.s. Partial-connection}

\author{\IEEEauthorblockN{Zheda Li, Shengqian Han, \textit{Member, IEEE}, Andreas F. Molisch, \textit{Fellow,~IEEE}}
\thanks{Z. Li is with DOCOMO Innovations, Inc., Palo Alto, California 94304, USA (email: zli@docomoinnovations.com). S. Han is with the School of Electronics and Information Engineering, Beihang University, Beijing 100191, China (e-mail:sqhan@buaa.edu.cn). A. F. Molisch is with the Department of Electrical Engineering, University of Southern California, Los Angeles, California 90089, USA (e-mail: molisch@usc.edu).}}
	
\maketitle

\begin{abstract}
In this article we compare the performance of two typical hyrbid beamforming structures for multiuser massive MIMO systems, i.e., the full- and partial-connection structures. Under the assumption of small angular spread for mmWave channels, given the analog precoder formed towards users and the zero-forcing digital precoder, we develop an explicit upper bound for the analog-and-digital-precoded channel gain of users, based on which the relationship between the two structures is investigated. The analysis results show that the full-connection structure is not always better than the partial-connection structure, and the regimes suitable for each structure are revealed. Simulations are conducted to validate the analysis results.
\end{abstract}


\section{Introduction}
To explore the large available bandwidth in the millimeter (mm)-wave bands~\cite{FCC2016} for the fifth generation (5G) cellular systems, massive multiple-input multiple-output (MIMO) technology is essential since the high
free-space pathloss at those frequencies necessitates large array gains to obtain sufficient signal-to-noise ratio (SNR)~\cite{Marzetta2010,Molisch2017}. By employing dozens or hundreds of antenna elements at the base station (BS), massive MIMO is able to significantly increase the spectral efficiency and simplify signal processing~\cite{Marzetta2010,Molisch2017,Fredrik2013}. However, implementation of such systems suffers from the prohibitive cost and high energy consumption to build a complete radio frequency (RF) chain for each antenna element, especially at mm-wave frequencies. A promising solution to these problems lies in the concept of hybrid beamforming (HB) transceivers, which uses a combination of analog beamformers in the RF domain, together with a smaller number of RF chains. This concept was first proposed in~\cite{Zhang2005} and~\cite{Sundarshan2006}. While formulated originally for MIMO with arbitrary number of antenna elements, the approach is applicable in particular to massive MIMO, and in that context interest in hybrid transceivers has surged over the past years. 

\begin{figure}
	\centering
    \subfigure[]
        {
        \includegraphics[width=0.47\textwidth]{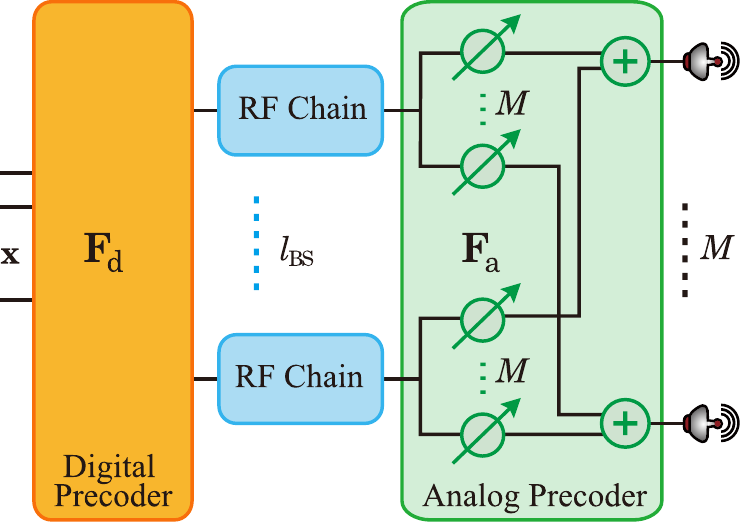}
        }
    \subfigure[]
        {
        \includegraphics[width=0.47\textwidth]{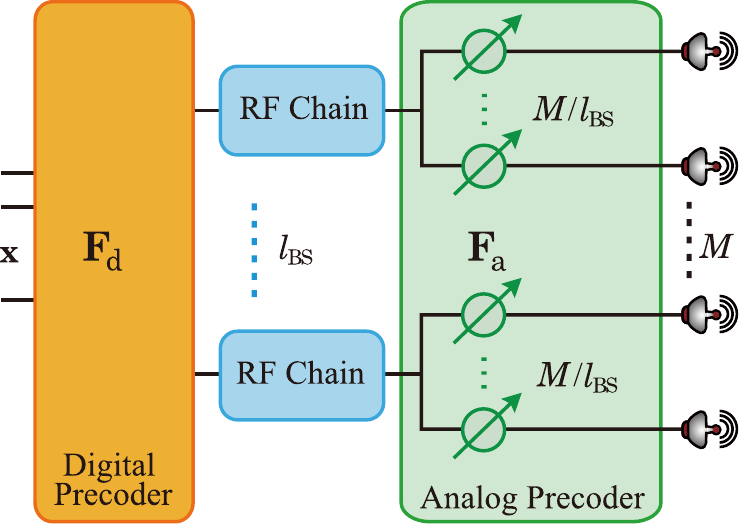}
        }
	\caption{Hybrid beamforming structures with (a) full-connection and (b) partial-connection.}\label{F:structure}
\end{figure}

Two typical HB structures at the BS, namely full- and partial-connection, are illustrated in Fig.~\ref{F:structure}, where the downlink transmission is assumed while the structure for uplink transmission is
similar. The difference between the two structures lies in the phase-shifter network for the analog beamformer. For a full-connection HB structure as in Fig.~\ref{F:structure}(a), each analog precoder output can be
a linear combination of all RF signals. Complexity reduction can be achieved when
each RF chain can be connected only to a subset of antenna elements, as in Fig.~\ref{F:structure}(b).

The majority of existing works on hybrid beamformer design considered the full-connection structure, e.g.,~\cite{Zhang2005,Ayach2014,Alkhateeb2015,Sohrab2016,Bogale2016,Ni2015,Sundarshan2006,Ansuman2013,Ansuman2014,Li2017,Li2018a,Li2018,Ioushua2019}, and some considered the partial-connection structure, e.g.,~\cite{partial-Gao2016,parital-Yu2016,partial-Park2017,Majidzadeh2018,Dsouza2018,Zhang2018,Ioushua2019}.
It is commonly accepted that the partial-connection structure enjoys the reduced cost and complexity compared to the full-connection structure with the penalty of certain data rate loss. This is true when the hybrid beamformers for the two structures are optimized under the identical objective function and constraints, as demonstrated in e.g.,~\cite{partial-Gao2016,parital-Yu2016,partial-Park2017,Majidzadeh2018,Dsouza2018,Zhang2018,Ioushua2019}. The reason is obvious, i.e., the analog beamformer for the full-connection structure has larger optimization freedoms (i.e., more variables to be optimized) than that for the partial-connection structure. In practical applications, however, to reduce the processing complexity and signaling overhead for CSI acquisition, suboptimal analog beamformers are usually applied. For example, in the 5G New Radio systems, the analog precoder is a combination of Discrete Fourier Transform (DFT) vectors selected from a predefined codebook by user equipments (UEs)~\cite{38.214}. In such scenarios, it is unclear whether the partial-connection structure always sacrifices the performance or not compared to the full-connection structure.

In this article, we compare the downlink performance of two HB structures for multiuser massive MIMO systems. By assuming a small angular spread for the mm-Wave channels, which leads to the analog precoder at the BS formed towards UEs, we derive an explicit upper bound for the analog-and-digital-precoded channel gain, where the zero-forcing digital precoder is employed. Based on the result, the gain of the full-connection structure over the partial-connection structure is analyzed. The analysis results show that the full-connection structure is not always better, and the regimes suitable for the two structures are revealed.

\section{System model}

\subsection{Signal Model}
Consider the downlink transmission in a single cell with $K$ single-antenna UEs and a BS equipped with $M$ antennas and $l_{\text{BS}}$ RF chains. The baseband digital precoder $\mathbf{F}_\text{d}$ processes the data streams $\mathbf{x}\sim\mathcal{CN}(\mathbf{0}, \mathbf{I})$ to produce $l_{\text{BS}}$ outputs,
which are upconverted to RF and mapped via an analog precoder $\mathbf{F}_\text{a}$ to $M$ antenna elements
for transmission. We consider both the full-connection and partial-connection HB structures for the BS, as shown in Fig.~\ref{F:structure}.

The received signal at the $\text{UE}_i$ is
\begin{align}
y_i&=\mathbf{h}_{i}^\dag\mathbf{F}_\text{a}\mathbf{F}_\text{d}\sqrt{p}\mathbf{x}+n_i
=\mathbf{h}_{i}^\dag\mathbf{F}_\text{a}\mathbf{f}_{\text{d,}i}\sqrt{p}x_i+\sum_{j\neq i} \mathbf{h}_i^\dag\mathbf{F}_\text{a}\mathbf{f}_{\text{d,}j}\sqrt{p}x_j+n_i, 
\end{align}
where $\mathbf{h}_i\in\mathbb{C}^{M\times 1}$ is the downlink channel of $\text{UE}_i,  \forall i$, $\mathbf{F}_{\text{a}}\in\mathbb{C}^{M\times l_{\text{BS}}}$ is the analog precoder, $\mathbf{F}_\text{d}=[\mathbf{f}_{\text{d,}1},...,\mathbf{f}_{\text{d,}K}]\in\mathbb{C}^{l_{\text{BS}}\times K}$ indicates the digital precoder, $\mathbf{f}_{\text{d,}i}$ is the digital precoder for $\text{UE}_i$, $n_i\sim\mathcal{CN}(0,\delta_n^2)$ is the complex Gaussian noise at $\text{UE}_i$, and
$p = \frac{P_t}{\|\mathbf{F}_\text{a}\mathbf{F}_\text{d}\|_F^2}$ scales the transmitted signal to satisfy the transmit power constraint of the BS with $P_t$ denoting the total transmit power of the BS and $\|\cdot\|_F$ denoting the Frobenius norm.

Define the analog-preocded effective channel as $\bar{\mathbf{h}}_i\triangleq \mathbf{F}_\text{a}^\dag\mathbf{h}_i, \forall i$, which has a low dimension and can be estimated with low overhead. Assume that the BS knows $\bar{\mathbf{h}}_i, \forall i$, and employs the zero-forcing digital precoding. Then, the digital precoder to serve $\text{UE}_i$ can be expressed as
\begin{align}
\mathbf{f}_{\text{d,}i}=\frac{\Pi(\tilde{\mathbf{H}}_i)\bar{\mathbf{h}}_i}{\|\Pi(\tilde{\mathbf{H}}_i)\bar{\mathbf{h}}_i\|},
\end{align}
where $\tilde{\mathbf{H}}_i\triangleq [\bar{\mathbf{h}}_1,...,\bar{\mathbf{h}}_{i-1},\bar{\mathbf{h}}_{i+1},...,\bar{\mathbf{h}}_K]$, and $\Pi(\tilde{\mathbf{H}}_i)\triangleq \mathbf{I}_{l_\text{BS}}-\tilde{\mathbf{H}}_i(\tilde{\mathbf{H}}_i^\dag \tilde{\mathbf{H}}_i)^{-1}\tilde{\mathbf{H}}_i^\dag$ indicates the null space of $\tilde{\mathbf{H}}_i$.
Then, the received signal at $\text{UE}_i$ can be represented as
\begin{align}
y_i=&\bar{\mathbf{h}}_i^\dag\mathbf{f}_{\text{d,}i}\sqrt{p}x_i+\sum_{j\neq i}^K \bar{\mathbf{h}}_i^\dag\mathbf{f}_{\text{d,}j}\sqrt{p}x_j+n_i = \bar{\mathbf{h}}_i^\dag\mathbf{f}_{\text{d,}i}\sqrt{p}x_i+n_i.
\end{align}

We define the analog-digital-precoded effective channel gain of $\text{UE}_i$ as
\begin{align} \label{E:effective-gain}
h_i\triangleq &|\bar{\mathbf{h}}_i^\dag\mathbf{f}_{\text{d,}i}|^2 = \frac{|\bar{\mathbf{h}}_i^\dag\Pi(\tilde{\mathbf{H}}_i)\bar{\mathbf{h}}_i|^2}
{\|\Pi(\tilde{\mathbf{H}}_i)\bar{\mathbf{h}}_i\|^2}=
\bar{\mathbf{h}}_i^\dag\Pi(\tilde{\mathbf{H}}_i)\bar{\mathbf{h}}_i=
\bar{\mathbf{h}}_i^\dag\tilde{\mathbf{U}}_i\tilde{\mathbf{U}}_i^\dag\bar{\mathbf{h}}_i
=\|\tilde{\mathbf{U}}_i^\dag\bar{\mathbf{h}}_i\|^2,
\end{align}
where the projection matrix $\Pi(\tilde{\mathbf{H}}_i)$ is decomposed as $\Pi(\tilde{\mathbf{H}}_i)=\tilde{\mathbf{U}}_i\tilde{\mathbf{U}}_i^\dag$, and $\tilde{\mathbf{U}}_i\in\mathbb{C}^{l_\text{BS}\times (l_\text{BS}-K+1)}$ is semi-unitary.


\subsection{Channel Model}
We consider the following propagation channel description:
\begin{align}
\mathbf{h}_i=\frac{1}{\sqrt{\alpha_i}}\sum_{l=1}^{L_i}a_{il}\mathbf{b}_M(\theta_{il}),
\label{Eq.channel_prop}
\end{align}
where $L_i$ is the number of scatterers between the BS and $\text{UE}_i$, $a_{il}\sim\mathcal{CN}(0,\delta_{il}^2)$ indicates the fading of $l$-th MPC to $\text{UE}_i$, $\alpha_i$ indicates the overall large-scale loss, and $\mathbf{b}_M(\theta_{il})$ is the $M$-by-$1$ steering vector with respect to angle of departure (AOD) $\theta_{il}$, $\forall i, l$. We consider a one-ring cluster model, where every subpath has the equal power, i.e., $\delta_{il}^2=\frac{1}{L_i}$, and the AODs of $\text{UE}_i$ follow uniform distribution $\mathcal{U}(\theta_i-\Delta_i,\theta_i+\Delta_i)$ with $\theta_i$ denoting the center angle and $\Delta_i$ denoting the angle spread. For the uniform linear array (ULA), the $m$-th entry of $\mathbf{h}_i$ can be represented as
\begin{align}
\mathbf{h}_i(m)=\frac{1}{\sqrt{\alpha_i}}\sum_{l=1}^{L_i}a_{il}e^{-j2\pi(m-1)d\cos(\theta_i+\tilde{\theta}_{il})},
\label{Eq.channelEntry}
\end{align}
where $\tilde{\theta}_{il}\sim \mathcal{U}(-\Delta_i,\Delta_i)$.

Assuming a small AOD spread for the mmWave channels, i.e., $\Delta_i\approx 0$, we can approximate $\cos(\theta_i+\tilde{\theta}_{il})$ in \eqref{Eq.channelEntry} as
\begin{align}
&\cos(\theta_i+\tilde{\theta}_{il}){=}\cos(\theta_i)\cos(\tilde{\theta}_{il})-\sin(\theta_i)
\sin(\tilde{\theta}_{il}){\approx}\cos\theta_i-\sin\theta_i\tilde{\theta}_{il},
\label{Eq.cosSimplify}
\end{align}
where the approximation holds when $\tilde{\theta}_{il}$ approaches zero, $\forall i,l$. Substitute (\ref{Eq.cosSimplify}) into (\ref{Eq.channelEntry}), we can approximate $\mathbf{h}_i(m)$ by $\tilde{\mathbf{h}}_i(m)$ as
\begin{align}
\tilde{\mathbf{h}}_i(m)=&\frac{1}{\sqrt{\alpha_i}}\sum_{l=1}^{L_i}a_{il}e^{-j2\pi(m-1)d(\cos\theta_i-\sin\theta_i
\tilde{\theta}_{il})},
\label{Eq.channelApprox}
\end{align}
and the steering vector ${\mathbf{b}}_M(\theta_{il})\triangleq [1,e^{-j2\pi d\cos\theta_{il}},...,e^{-j2\pi(M-1)d\cos\theta_{il}}]^T$ can be approximated by $\tilde{\mathbf{b}}_M(\theta_i,\tilde{\theta}_{il})$ as
\begin{align}
    \tilde{\mathbf{b}}_M(\theta_i,\tilde{\theta}_{il})\triangleq [1,e^{-j2\pi d(\cos\theta_i-\sin\theta_i\tilde{\theta}_{il})},...,e^{-j2\pi(M-1)d(\cos\theta_i-\sin\theta_i\tilde{\theta}_{il})}]^T.
\end{align}

\subsection{Analog Precoder}
Under the scenarios with small AOD spread $\Delta_i\approx 0$, we consider that the analog precoder is formed towards the central angle of UEs.
For the full-connection structure, the analog precoder, denoted by $\mathbf{F}_\text{aF}$, consists of the normalized steering vectors of UEs, which is
\begin{align}
&\mathbf{F}_\text{aF}=[\mathbf{f}_{\text{aF,}1},...,\mathbf{f}_{\text{aF,}K}] \ \text{with} \  \mathbf{f}_{\text{aF,}k}=\frac{1}{\sqrt{M}}\mathbf{b}_M(\theta_k),\forall k.\label{E:Analog-full}
\end{align}

For the partial-connection structure, suppose that each UE is served by a subarray as commonly considered in the literature, e.g., \cite{partial-Gao2016,Majidzadeh2018}, where full multiplexing is assumed to fully exploit all RF and antenna elements. In this paper, we will consider both full and partial multiplexing cases with $K\leq l_{\text{BS}}$. If $K < l_{\text{BS}}$, then there will be $l_{\text{BS}}-K$ unused RF chains and $(l_{\text{BS}}-K)M_{\text{P}}$ unused antenna elements, where $M_{\text{P}} = M/l_{\text{BS}}$ is the number of antenna elements per subarray.
The analog precoder, denoted by $\mathbf{F}_\text{aP}$, can be expressed as
\begin{align}
&\mathbf{F}_\text{aP}=\left[\diag{([\mathbf{f}_{\text{aP},k}^T]_{k=1}^K)} \ \mathbf{0}\right]^T  \ \text{with} \  \mathbf{f}_{\text{aP},k}=\frac{1}{\sqrt{M_\text{P}}}\mathbf{b}_{M_\text{P}}(\theta_k), \forall k,\label{E:Analog-part}
\end{align}
where $\diag(\cdot)$ generates the block diagonal matrix. It can found that $\mathbf{F}_\text{aP}$ has $(l_{\text{BS}}-K)M_{\text{P}}$ rows of zeros, corresponding to unused antenna elements.

\section{Performance Analysis for Different HB Structures}
In this section we conduct theoretical analyses for the performance of the hybrid precoders under the full- and partial-connection HB structures. We first derive an upper bound for the sum rate under the zero-forcing digital precoder, then given the analog precoder as the steering vectors toward the central angles of UEs, we compare the performance of the two HB structures.

\subsection{Upper Bound of Average Sum Rate }
With the zero-forcing digital precoder, the effective channel gain $h_i$ defined in~\eqref{E:effective-gain} can be bounded as follows.

\emph{\textbf{Lemma 1:} For a given UE set and analog precoder, the effective channel gain is upper bounded by}
\begin{align}
h_i\leq \|\bar{\mathbf{h}}_i\|^2- \frac{|\bar{\mathbf{h}}_i^\dag\bar{\mathbf{h}}_j|^2}{\|\bar{\mathbf{h}}_j\|^2}, \forall j\neq i.
\end{align}
\begin{IEEEproof}
 The Lemma follows from the fact that the orthogonal projection of a vector to a subspace is not larger than that to a vector involved in the subspace. A detailed proof is given in Appendix~\ref{A:proof-lemma1}.
\end{IEEEproof}

According to Lemma 1, we obtain the following inequality
\begin{align}
h_i\leq \|\bar{\mathbf{h}}_i\|^2-\frac{1}{K-1}\sum_{j\neq i}^K\frac{|\bar{\mathbf{h}}_j^\dag\bar{\mathbf{h}}_i|^2}{\|\bar{\mathbf{h}}_j\|^2}.
\end{align}

Then, we can acquire an upper bound for the achievable rate
\begin{align}
R_\text{sum}=\sum_{k=1}^K\log(1+\frac{h_kp}{\delta_\text{n}^2})\leq \sum_{k=1}^K\log(1+\frac{\|\bar{\mathbf{h}}_k\|^2-\frac{1}{K-1}\sum_{j\neq k}^K\frac{|\bar{\mathbf{h}}_j^\dag\bar{\mathbf{h}}_k|^2}{\|\bar{\mathbf{h}}_j\|^2}}{\delta_\text{n}^2}p)
\end{align}

Taking the expectation at both sides, we can get an upper bound of the average sum rate by Jensen's inequality as
\begin{align}
\mathbb{E}[R_\text{sum}]\leq &\sum_{k=1}^K\mathbb{E}\left[\log\left(1+\frac{\|\bar{\mathbf{h}}_k\|^2-\frac{1}{K-1}\sum_{j\neq k}^K\frac{|\bar{\mathbf{h}}_j^\dag\bar{\mathbf{h}}_k|^2}{\|\bar{\mathbf{h}}_j\|^2}}{\delta_\text{n}^2}p\right)\right]\nonumber\\
\leq & \sum_{k=1}^K \log\left(1+\frac{\mathbb{E}\left[\|\bar{\mathbf{h}}_k\|^2\right]-\frac{1}{K-1}\sum_{j\neq k}^K \mathbb{E}\left[\frac{|\bar{\mathbf{h}}_j^\dag\bar{\mathbf{h}}_k|^2}{\|\bar{\mathbf{h}}_j\|^2} \right]}{\delta_\text{n}^2}p\right)\triangleq \bar{R}_{\text{sum}}^{\text{ub}}.
\label{Eq.sumRateUpper}
\end{align}


Next, we develop analytic expressions of $\mathbb{E}[\|\bar{\mathbf{h}}_k\|^2]$ and $\mathbb{E}[\frac{|\bar{\mathbf{h}}_j^\dag\bar{\mathbf{h}}_k|^2}{\|\bar{\mathbf{h}}_j\|^2}]$ in (\ref{Eq.sumRateUpper}).

With (\ref{Eq.channelApprox}), $\mathbb{E}[|\bar{\mathbf{h}}_j^\dag\bar{\mathbf{h}}_k|^2]$ can be expanded as
\begin{align} \label{E:cross}
\mathbb{E}[|\bar{\mathbf{h}}_j^\dag\bar{\mathbf{h}}_k|^2]
\approx&\frac{1}{\alpha_j\alpha_k}\mathbb{E}[|\sum_{x=1}^{P_j}\sum_{y=1}^{P} a_{jx}^\ast a_{ky}\tilde{\mathbf{b}}_M^\dag(\theta_j,\tilde{\theta}_{jx})\mathbf{F}_\text{a}\mathbf{F}_\text{a}^\dag\tilde{\mathbf{b}}_M(\theta_k,\tilde{\theta}_{ky})|^2]\nonumber\\
=&\frac{1}{\alpha_j\alpha_k}\sum_{x_1=1}^{P_j}\sum_{y_1=1}^{P}\sum_{x_2=1}^{P_j}\sum_{y_2=1}^{P}\mathbb{E}[ a_{jx_1}^\ast a_{jx_2}]\mathbb{E}[ a_{ky_1} a_{ky_2}^\ast]\cdot\nonumber\\
&\qquad\mathbb{E}[\tilde{\mathbf{b}}_M^\dag(\theta_j,\tilde{\theta}_{jx_1})\mathbf{F}_\text{a}\mathbf{F}_\text{a}^\dag\tilde{\mathbf{b}}_M(\theta_k,\tilde{\theta}_{ky_1})\tilde{\mathbf{b}}_M^\dag(\theta_k,\tilde{\theta}_{ky_2})\mathbf{F}_\text{a}\mathbf{F}_\text{a}^\dag\tilde{\mathbf{b}}_M(\theta_j,\tilde{\theta}_{jx_2})]\nonumber\\
=&\frac{1}{\alpha_j\alpha_k}\tr(\mathbf{F}_\text{a}^\dag\mathbb{E}[\tilde{\mathbf{b}}_M(\theta_k,\tilde{\theta}_{ky_1})\tilde{\mathbf{b}}_M^\dag(\theta_k,\tilde{\theta}_{ky_1})]\mathbf{F}_\text{a}\mathbf{F}_\text{a}^\dag\mathbb{E}[\tilde{\mathbf{b}}_M(\theta_j,\tilde{\theta}_{jx_1})\tilde{\mathbf{b}}_M^\dag(\theta_j,\tilde{\theta}_{jx_1})]\mathbf{F}_\text{a})\nonumber\\
\triangleq&\frac{1}{\alpha_j\alpha_k}\tr(\mathbf{F}_\text{a}^\dag\tilde{\mathbf{K}}_k\mathbf{F}_\text{a}\mathbf{F}_\text{a}^\dag\tilde{\mathbf{K}}_j\mathbf{F}_\text{a}),
\end{align}
where we define $\tilde{\mathbf{K}}_j\triangleq \mathbb{E}[\tilde{\mathbf{b}}_M(\theta_j,\tilde{\theta}_{jx_1})\tilde{\mathbf{b}}_M^\dag(\theta_j,\tilde{\theta}_{jx_1})]$ with the $(m,n)$-th entry
\begin{align} \label{E:self}
\tilde{\mathbf{K}}_j(m,n)&=\mathbb{E}[e^{-j2\pi(m-1)d(\cos\theta_j-\sin\theta_j\tilde{\theta}_{jx})}e^{j2\pi(n-1)d(\cos\theta_j-\sin\theta_j\tilde{\theta}_{jx})}]\nonumber\\
&=e^{j2\pi(n-m)d\cos\theta_j}\mathbb{E}[e^{j2\pi(m-n)d\sin\theta_j\tilde{\theta}_{jx}}]\nonumber\\
&=e^{j2\pi(n-m)d\cos\theta_j}\frac{\sin(2\pi d(m-n)\sin\theta_j\Delta_j)}{2\pi d(m-n)\sin\theta_j\Delta_j},
\end{align}
where the final equality is obtained by considering $\tilde{\theta}_{jx}\sim\mathcal{U}(-\Delta_i,\Delta_i)$.

The term $\mathbb{E}[\|\bar{\mathbf{h}}_i\|^2]$ can be derived as
\begin{align}
\mathbb{E}[\|\bar{\mathbf{h}}_i\|^2]&\approx\frac{1}{\alpha_i}\mathbb{E}[\sum_{x=1}^{L_i}\sum_{y=1}^{L_i} a_{ix}^\ast a_{iy}\tilde{\mathbf{b}}_M^\dag(\theta_i,\tilde{\theta}_{ix})\mathbf{F}_\text{a}\mathbf{F}_\text{a}^\dag\tilde{\mathbf{b}}_M(\theta_i,\tilde{\theta}_{iy})]\nonumber\\
&=\frac{1}{\alpha_i}\mathbb{E}[\tilde{\mathbf{b}}_M^\dag(\theta_i,\tilde{\theta}_{ix})\mathbf{F}_\text{a}\mathbf{F}_\text{a}^\dag\tilde{\mathbf{b}}_M(\theta_i,\tilde{\theta}_{ix})]
=\frac{1}{\alpha_i}\tr(\mathbf{F}_\text{a}^\dag\tilde{\mathbf{K}}_i\mathbf{F}_\text{a}).
\end{align}

With \eqref{E:cross} and \eqref{E:self}, the upper bound given in (\ref{Eq.sumRateUpper}) can be rewritten as
\begin{align}
 \bar{R}_{\text{sum}}^{\text{ub}}& \approx\sum_{k=1}^K \log(1+\frac{\tr(\mathbf{F}_\text{a}^\dag\tilde{\mathbf{K}}_k\mathbf{F}_\text{a})-\frac{1}{K-1}\sum_{j\neq k}^K\frac{\tr{(\mathbf{F}_\text{a}^\dag\tilde{\mathbf{K}}_k\mathbf{F}_\text{a}\mathbf{F}_\text{a}^\dag
\tilde{\mathbf{K}}_j\mathbf{F}_\text{a})}}{\tr(\mathbf{F}_\text{a}^\dag\tilde{\mathbf{K}}_j\mathbf{F}_\text{a})}}
{\alpha_k\delta_n^2}p)\nonumber\\
 &\triangleq\sum_{k=1}^K\log(1+\frac{g_kp}{\alpha_k\delta_n^2}),
 \label{Eq.upperBound}
\end{align}
where we define $g_k\triangleq \tr(\mathbf{F}_\text{a}^\dag\tilde{\mathbf{K}}_k\mathbf{F}_\text{a})-\frac{1}{K-1}\sum_{j\neq k}^K\frac{\tr{(\mathbf{F}_\text{a}^\dag\tilde{\mathbf{K}}_k\mathbf{F}_\text{a}\mathbf{F}_\text{a}^\dag
\tilde{\mathbf{K}}_j\mathbf{F}_\text{a})}}{\tr(\mathbf{F}_\text{a}^\dag\tilde{\mathbf{K}}_j\mathbf{F}_\text{a})}$, which determines the data rate of UE$_k$. 

To compare the performance under two HB structures, we need to obtain an explicit expression of $g_k$ as defined below \eqref{Eq.upperBound}. Under the scenarios with small AOD spread $\Delta_i\approx 0$,
%
%
%
we can obtain from \eqref{E:self} that
\begin{align}
\tilde{\mathbf{K}}_j(m,n)\approx e^{j2\pi(n-m)d\cos\theta_j}, \ \ \tilde{\mathbf{K}}_j\approx \mathbf{b}_M(\theta_j)\mathbf{b}_M^\dag(\theta_j).
\end{align}
Then, we can further simplify (\ref{Eq.upperBound}) by considering that
\begin{align} \label{E:simplify}
\tr(\mathbf{F}_\text{a}^\dag\tilde{\mathbf{K}}_k\mathbf{F}_\text{a})
=\|\mathbf{F}_\text{a}^\dag\mathbf{b}_M(\theta_k)\|^2,
\ \
\tr{(\mathbf{F}_\text{a}^\dag\tilde{\mathbf{K}}_k\mathbf{F}_\text{a}\mathbf{F}_\text{a}^\dag\tilde{\mathbf{K}}_j
\mathbf{F}_\text{a})}
=|\mathbf{b}_M^\dag(\theta_k)\mathbf{F}_\text{a}\mathbf{F}_\text{a}^\dag\mathbf{b}_M(\theta_j)|^2.
\end{align}


We first consider the full-connection structure by setting $\mathbf{F}_\text{a}=\mathbf{F}_\text{aF}$.
Upon substituting \eqref{E:Analog-full} into \eqref{E:simplify}, we can update $\tr(\mathbf{F}_\text{a}^\dag\tilde{\mathbf{K}}_k\mathbf{F}_\text{a})$ as
\begin{align}
&\tr(\mathbf{F}_\text{aF}^\dag\tilde{\mathbf{K}}_k\mathbf{F}_\text{aF})=\frac{1}{M}\sum_{j=1}^K\left|\mathbf{b}_M^\dag(\theta_j)\mathbf{b}_M(\theta_k)\right|^2\nonumber\\
&=\frac{1}{M}\sum_{j=1}^K\left|\sum_{n=1}^M e^{j2\pi d(n-1)(\cos\theta_j-\cos\theta_k)}\right|^2
=\frac{1}{M}\sum_{j=1}^K\left|\frac{\sin(\pi dM(\cos\theta_j-\cos\theta_k))}{\sin(\pi d(\cos\theta_j-\cos\theta_k))}\right|^2, \label{E:F-1}
\end{align}
and update $\tr{(\mathbf{F}_\text{a}^\dag\tilde{\mathbf{K}}_k\mathbf{F}_\text{a}\mathbf{F}_\text{a}^\dag\tilde{\mathbf{K}}_j
\mathbf{F}_\text{a})}$ as
\begin{align}
&\tr{(\mathbf{F}_\text{aF}^\dag\tilde{\mathbf{K}}_k\mathbf{F}_\text{aF}\mathbf{F}_\text{aF}^\dag
\tilde{\mathbf{K}}_j\mathbf{F}_\text{aF})}=\left|\frac{1}{M}\mathbf{b}_M^\dag(\theta_k)\sum_{i=1}^K\mathbf{b}_M(\theta_i)\mathbf{b}_M^\dag(\theta_i)\mathbf{b}_M(\theta_j)\right|^2\nonumber\\
&=\frac{1}{M^2}\left|\sum_{i=1}^K\sum_{n=1}^Me^{j2\pi d(n-1)(\cos\theta_k-\cos\theta_i)}\sum_{m=1}^Me^{j2\pi d(m-1)(\cos\theta_i-\cos\theta_j)}\right|^2\nonumber\\
&=\frac{1}{M^2}\left|\sum_{i=1}^K{e^{j\pi d(M-1)(\cos\theta_k-\cos\theta_j)}}\frac{\sin(\pi M d(\cos\theta_k-\cos\theta_i))}{\sin(\pi d(\cos\theta_k-\cos\theta_i))}\frac{\sin(\pi M d(\cos\theta_i-\cos\theta_j))}{\sin(\pi d(\cos\theta_i-\cos\theta_j))}\right|^2\nonumber\\
&=\frac{1}{M^2}\left|\sum_{i=1}^K\frac{\sin(\pi M d(\cos\theta_k-\cos\theta_i))}{\sin(\pi d(\cos\theta_k-\cos\theta_i))}\frac{\sin(\pi M d(\cos\theta_i-\cos\theta_j))}{\sin(\pi d(\cos\theta_i-\cos\theta_j))}\right|^2. \label{E:F-2}
\end{align}

With \eqref{E:F-1} and \eqref{E:F-2}, we can obtain the expression of $g_k$ as defined below \eqref{Eq.upperBound} for the full-connected structure, denoted by $g_{\text{F},k}$, as
\begin{align} \label{E:g-full}
&g_{\text{F},k}=\frac{1}{M}\sum_{i=1}^K\left|\frac{\sin(\pi dM(\cos\theta_i-\cos\theta_k))}{\sin(\pi d(\cos\theta_i-\cos\theta_k))}\right|^2-\nonumber\\
&\qquad \qquad \qquad \frac{1}{K-1}\sum_{j\neq k}^K \frac{\left|\sum_{i=1}^K\frac{\sin(\pi M d(\cos\theta_k-\cos\theta_i))}{\sin(\pi d(\cos\theta_k-\cos\theta_i))}\frac{\sin(\pi M d(\cos\theta_i-\cos\theta_j))}{\sin(\pi d(\cos\theta_i-\cos\theta_j))}\right|^2}{M\sum_{i=1}^K\left|\frac{\sin(\pi dM(\cos\theta_i-\cos\theta_j))}{\sin(\pi d(\cos\theta_i-\cos\theta_j))}\right|^2}.
\end{align}

We proceed to consider the partial-connection structure by setting $\mathbf{F}_\text{a}=\mathbf{F}_\text{aP}$. Upon substituting \eqref{E:Analog-part} into \eqref{E:simplify}, we can update $\tr(\mathbf{F}_\text{a}^\dag\tilde{\mathbf{K}}_k\mathbf{F}_\text{a})$ as
\begin{align}
\tr(\mathbf{F}_{\text{aP}}^\dag\tilde{\mathbf{K}}_k\mathbf{F}_\text{aP})&=\sum_{i=1}^K\left|\frac{1}{\sqrt{M_\text{P}}}e^{-j2\pi d(i-1)M_\text{P}\cos\theta_k}e^{j\pi d (M_\text{P}-1)(\cos\theta_i-\cos\theta_k)}\frac{\sin(\pi d M_\text{P}(\cos\theta_i-\cos\theta_k))}{\sin(\pi d (\cos\theta_i-\cos\theta_k))}\right|^2\nonumber\\
&=\frac{1}{M_\text{P}}\sum_{i=1}^K \left|\frac{\sin(\pi d M_\text{P}(\cos\theta_i-\cos\theta_k))}{\sin(\pi d (\cos\theta_i-\cos\theta_k))}\right|^2, \label{E:P-1}
\end{align}
and update $\tr{(\mathbf{F}_\text{a}^\dag\tilde{\mathbf{K}}_k\mathbf{F}_\text{a}\mathbf{F}_\text{a}^\dag\tilde{\mathbf{K}}_j
\mathbf{F}_\text{a})}$ as
\begin{align}
&\tr{(\mathbf{F}_\text{aP}^\dag\tilde{\mathbf{K}}_k\mathbf{F}_\text{aP}\mathbf{F}_\text{aP}^\dag\tilde{\mathbf{K}}_j
\mathbf{F}_\text{aP})}
=\left|\sum_{i=1}^K\frac{1}{\sqrt{M_\text{P}}}e^{j2\pi d(i-1)M_\text{P}\cos\theta_k}\sum_{n=1}^{M_\text{P}}e^{j2\pi d (n-1)(\cos\theta_k-\cos\theta_i)}\cdot\right.\nonumber\\
&\qquad\qquad\qquad\qquad\qquad\qquad\left.\frac{1}{\sqrt{M_\text{P}}}e^{-j2\pi d(i-1)M_\text{P}\cos\theta_j}\sum_{m=1}^{M_\text{P}}e^{j2\pi d (m-1)(\cos\theta_i-\cos\theta_j)}\right|^2\nonumber\\
&=\frac{1}{M_\text{P}^2}\left|\sum_{i=1}^Ke^{j2\pi d(i-1)M_\text{P}(\cos\theta_k-\cos\theta_j)}\frac{\sin(\pi d M_\text{P}(\cos\theta_k-\cos\theta_i))}{\sin(\pi d (\cos\theta_k-\cos\theta_i))}\frac{\sin(\pi d M_\text{P}(\cos\theta_j-\cos\theta_i))}{\sin(\pi d (\cos\theta_j-\cos\theta_i))}\right|^2.
\end{align}

Consequently, we can obtain the expression of $g_k$ for the partial-connected structure, denoted by $g_{\text{P},k}$, as
\begin{align}\label{E:g-part}
&g_{\text{P},k}=\frac{1}{M_\text{P}}\sum_{i=1}^K\left|\frac{\sin(\pi dM_\text{P}(\cos\theta_i-\cos\theta_k))}{\sin(\pi d(\cos\theta_i-\cos\theta_k))}\right|^2-\nonumber\\
&\qquad  \frac{1}{K-1}\sum_{j\neq k}^K \frac{\left|\sum_{i=1}^Ke^{j2\pi d(i-1)M_\text{P}(\cos\theta_k-\cos\theta_j)}\frac{\sin(\pi M_\text{P} d(\cos\theta_k-\cos\theta_i))}{\sin(\pi d(\cos\theta_k-\cos\theta_i))}\frac{\sin(\pi M_\text{P} d(\cos\theta_i-\cos\theta_j))}{\sin(\pi d(\cos\theta_i-\cos\theta_j))}\right|^2}{M_\text{P}\sum_{i=1}^K\left|\frac{\sin(\pi dM_\text{P}(\cos\theta_i-\cos\theta_j))}{\sin(\pi d(\cos\theta_i-\cos\theta_j))}\right|^2}.
\end{align}

\subsection{Structure Comparison in Special Cases}
Based on the obtained expressions of $g_{\text{F},k}$ and $g_{\text{P},k}$, we compare the two HB structures by considering some special cases in this subsection, and will consider the general case in the next subsection.

\subsubsection{\underline{Case 1: $M_\text{P}\rightarrow \infty$, $M\rightarrow \infty$, arbitrary $K$ UEs}}

When the number of antennas is sufficiently large, we can readily obtain from \eqref{E:g-full} and \eqref{E:g-part} that
\begin{align}
  &\lim_{M\rightarrow \infty} g_{\text{F},k}=M, \ \ \lim_{M_\text{P}\rightarrow\infty}g_{\text{P},k}=M_\text{P},  \ \
  \lim_{M_\text{P}\rightarrow\infty}\frac{g_{\text{F},k}}{g_{\text{P},k}}=l_\text{BS}.
\end{align}
The results show that when the antenna array has sufficiently high spatial resolution with a large number of antennas, the full-connection structure outperforms the partial-connection structure by the beamforming gain at the order of $l_\text{BS}$.

\subsubsection{\underline{Case 2: Arbitrary $M_\text{P}$ and $M$, $K=2$ UEs}}

Consider that the BS serves $K=2$ UEs. For an arbitrary UE, say UE$_1$, we can obtain from \eqref{E:g-full} and \eqref{E:g-part} that
\begin{align}
g_{\text{F},1}&=M+\frac{1}{M}Z_M^2(\beta_{12})-\frac{4MZ_M^2(\beta_{12})}{Z_M^2(\beta_{12})+M^2}=\frac{(M^2-Z_M^2(\beta_{12}))^2}{M(M^2+Z_M^2(\beta_{12}))},\nonumber\\
g_{\text{P},1}&=M_\text{P}+\frac{1}{M_\text{P}}Z_{M_\text{P}}^2(\beta_{12})-\frac{4M_\text{P}Z_{M_\text{P}}^2(\beta_{12})\cos^2(\pi dM_\text{P}\beta_{12})}{Z_{M_\text{P}}^2(\beta_{12})+M_\text{P}^2},\nonumber\\
&=\frac{(M_\text{P}^2+Z_{M_\text{P}}^2(\beta_{12}))^2-4M_\text{P}^2Z_{M_\text{P}}^2(\beta_{12})\cos^2(\pi dM_\text{P}\beta_{12})}{M_\text{P}(M_\text{P}^2+Z_{M_\text{P}}^2(\beta_{12}))}.
\end{align}
where we define the function $Z_M(x)\triangleq\frac{\sin(\pi d Mx)}{\sin(\pi d x)}$, and $\beta_{12}=\cos\theta_1-\cos\theta_2$ reflects the proximity of the two UEs.

Consequently, the ratio gap between $g_{\text{F},1}$ and $g_{\text{P},1}$ is
\begin{align} \label{E:ratio-1}
&\frac{g_{\text{F},1}}{g_{\text{P},1}}=\frac{(M^2-Z_M^2(\beta_{12}))^2}{(M_\text{P}^2+Z_{M_\text{P}}^2(\beta_{12}))^2-4M_\text{P}^2Z_{M_\text{P}}^2(\beta_{12})\cos^2(\pi d M_\text{P}\beta_{12})}\frac{M_\text{P}(M_\text{P}^2+Z_{M_\text{P}}^2(\beta_{12}))}{M(M^2+Z_M^2(\beta_{12}))}.
\end{align}

We next consider a toy example to simplify \eqref{E:ratio-1} to gain some insight, where the BS has $M=2$ antennas and $l_{\text{BS}}=2$ RF chains. In this case, each RF chain connects a single antenna in the partial-connection structure, i.e., $M_{\text{P}} = 1$. We can obtain that $Z_{M}(\beta_{12}) = \frac{\sin(2\pi d \beta_{12})}{\sin(\pi d \beta_{12})} = 2\cos(\pi d \beta_{12})$ and $Z_{M_\text{P}}(\beta_{12}) = 1$, with which \eqref{E:ratio-1} can be updated as
\begin{align}
    \frac{g_{\text{F},1}}{g_{\text{P},1}}
    =\frac{1-\cos^2(\pi d \beta_{12})}{ 1+\cos^2(\pi d \beta_{12})}.
\end{align}
It is obvious that $0\leq \frac{g_{\text{F},1}}{g_{\text{P},1}}\leq 1$ for the toy example, which indicates that the full-connection structure is worse than the partial-connection structure. Although the former is able to provide a larger beamforming gain, it also enhances the collinearity between the effective channels of the two UEs, making the overall performance worse.

\begin{figure}
\centering
\includegraphics[width=0.85\textwidth]{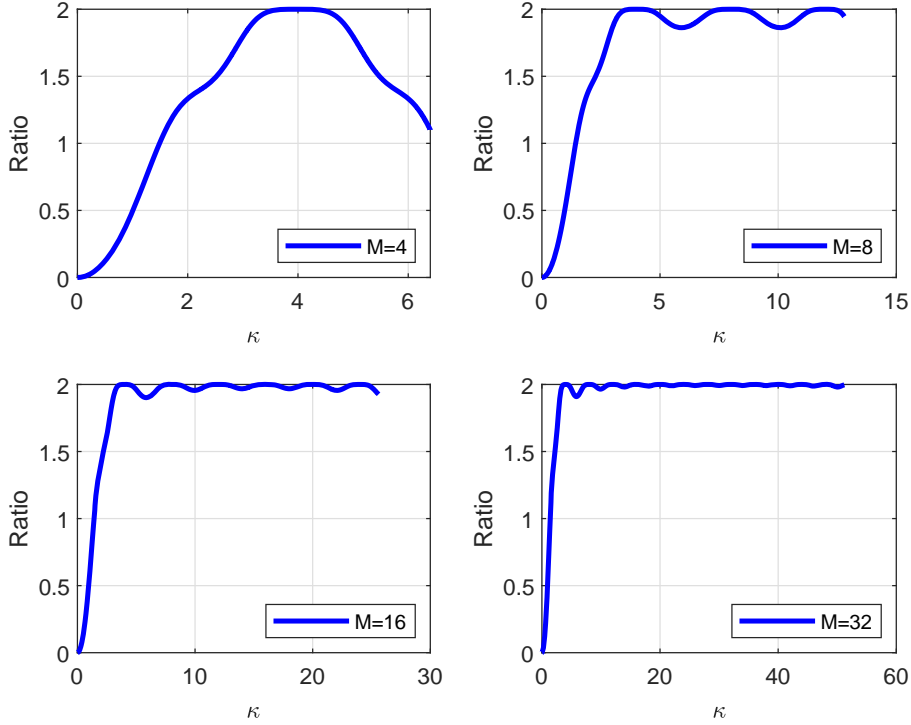}
\caption{Ratio $\frac{g_{\text{F},1}}{g_{\text{P},1}}$ v.s. $\kappa$ for $M=4, 8, 16$, and $32$.}
\label{Fig:ratio_vs_beta}
\end{figure}

For general cases, the value of $\frac{g_{\text{F},1}}{g_{\text{P},1}}$ in \eqref{E:ratio-1} depends on $M$, $l_{\text{BS}}$ and $\beta_{12}$. Numerically, we evaluate the value of $\frac{g_{\text{F},1}}{g_{\text{P},1}}$ for different $\kappa=\beta_{12}M$ in Fig. \ref{Fig:ratio_vs_beta}, where $l_{\text{BS}}=K=2$ is given. Herein, the x-axis $\kappa$ can be interpreted as the normalized angle separation between UEs by the angle resolution of the BS array, which is on the order of $\frac{1}{M}$.
We can see that the full-connection structure is not always better than the partial-connection structure as revealed by the toy example. The latter is better for small $\beta_{12}$, i.e., when the central angles to the two UEs, $\theta_1$ and $\theta_2$, are close. The regime of $\beta_{12}$ where the partial-connection structure is better decreases with the increase of $M$. When $M$ is very large, say 32, the ratio approaches a constant $l_\text{BS}$ as shown in Fig.~\ref{Fig:ratio_vs_beta}(d).

\subsection{Structure Comparison in General Cases}
The above special-case analysis implies that the full-connection structure may be interior to the partial-connection structure when the UEs have close central angles $\theta_i, \forall i$. In this subsection we focus on the regime where $\theta_i\approx \theta_j$, $\forall i\neq j$, aimed at investigating the condition of $\theta_i$ and $\theta_j$ when the full-connection structure performs better for a general case with arbitrary numbers of UEs $K\leq l_{\text{BS}}$ and antennas $M$.

Specifically, we assume that $\cos\theta_i-\cos\theta_j=\beta\approx 0$, $\forall i\neq j$. Then, we can simplify the expression of $g_{\text{F},k}$ in \eqref{E:g-full} with some regular manipulations as
\begin{align}
g_{\text{F},k}
&=\frac{[(2K-3)\zeta^2+2M\zeta+M^2](\zeta-M)^2}{M[M^2+(K-1)\zeta^2]}\approx \frac{M(\eta_\text{F}-1)^2[(2K-3)\eta_\text{F}^2+2\eta_\text{F}+1]}{1+(K-1)\eta_\text{F}^2}, \label{E:g-full-approx}
\end{align}
where $\zeta\triangleq \frac{\sin(\pi d \kappa)}{\sin(\pi d \frac{\kappa}{M})}$, $\kappa = \beta M$ is the normalized angle separation between UEs as defined in Fig.~\ref{Fig:ratio_vs_beta}, and the approximation follows from the first-order Taylor approximation of $\sin(\pi d \frac{\kappa}{M})$, which is accurate since $\frac{\kappa}{M} = \beta$ is small as we assumed, i.e., $\zeta\approx M\frac{\sin(\pi d\kappa)}{\pi d\kappa} \triangleq M\eta_\text{F}$. Herein, we define $\eta_\text{F} \triangleq \frac{\sin(\pi d\kappa)}{\pi d\kappa}$, and we can numerically obtain that its value ranges in $[-0.22,1]$.

Similarly, considering the approximation $\frac{\sin(\pi d \beta M_{\text{P}})}{\sin(\pi d \beta)} = \frac{\sin(\pi d \frac{\kappa}{l_{\text{BS}}})}{\sin(\pi d \frac{\kappa}{M})} \approx M\frac{\sin(\pi d \frac{\kappa}{l_{\text{BS}}})}{\pi d \kappa} =  M_{\text{P}}\frac{l_{\text{BS}}\sin(\pi d \frac{\kappa}{l_{\text{BS}}})}{\pi d \kappa} \triangleq M_{\text{P}}\eta_\text{P}$, we can simplify the expression of $g_{\text{P},k}$ in \eqref{E:g-part} with some regular manipulations as
\begin{align}
g_{\text{P},k} \approx M_\text{P}[1+(K-1)\eta_\text{P}^2]-\frac{M_\text{P}\eta_\text{P}^2}{K-1}
\frac{f_k(\frac{\kappa}{l_\text{BS}})}{1+(K-1)\eta_\text{P}^2},
\end{align}
where we define the function $f_k(\frac{\kappa}{l_\text{BS}})\triangleq \sum_{j\neq k}^K |e^{j2\pi d(k-1)\frac{\kappa}{l_\text{BS}}}+e^{j2\pi d(j-1)\frac{\kappa}{l_\text{BS}}}+\eta_\text{P}\sum_{i\neq j,k}e^{j2\pi d(i-1)\frac{\kappa}{l_\text{BS}}}|^2$ and $\eta_\text{P} = \frac{l_{\text{BS}}\sin(\pi d \frac{\kappa}{l_{\text{BS}}})}{\pi d \kappa}$.

Therefore, the ratio between $g_{\text{F},k}$ and $g_{\text{P},k}$ approximates
\begin{align} \label{E:approx-ratio}
\frac{g_{\text{F},k}}{g_{\text{P},k}}\approx\frac{l_\text{BS}(\eta_\text{F}-1)^2[(2K-3)
\eta_\text{F}^2+2\eta_\text{F}+1]}{[1+(K-1)\eta_\text{F}^2][1+(K-1)\eta_\text{P}^2-
\frac{\eta_\text{P}^2}{K-1}\frac{f_k(\frac{\kappa}{l_\text{BS}})}{1+(K-1)\eta_\text{P}^2}]}.
\end{align}

Since we are investigating the smallest separation between UEs to make the full-connection structure better, the interested normalized angle separation $\kappa = \beta M$ is generally small. Then, it is reasonable to assume $\beta M_{\text{P}} = \frac{\kappa}{l_{\text{BS}}}\approx 0$ since the number of RF chains $l_{\text{BS}}$ can be large for massive MIMO systems. With the approximation $\frac{\kappa}{l_{\text{BS}}}\approx 0$, we can obtain $\eta_\text{P}=\frac{l_{\text{BS}}\sin(\pi d \frac{\kappa}{l_{\text{BS}}})}{\pi d \kappa}\approx 1$ and then approximate $f_k(\frac{\kappa}{l_\text{BS}})$ by
\begin{align}
f_k(\frac{\kappa}{l_\text{BS}})&\approx \sum_{j\neq k}^K \left|\sum_{i}e^{j2\pi d(i-1)\frac{\kappa}{l_\text{BS}}}\right|^2
=\sum_{j\neq k}^K \left|\frac{1-e^{j2\pi dK\frac{\kappa}{l_\text{BS}}}}{1-e^{j2\pi d\frac{\kappa}{l_\text{BS}}}}\right|^2\nonumber\\
&=(K-1)\frac{\sin^2(\pi dK\frac{\kappa}{l_\text{BS}})}{\sin^2(\pi d\frac{\kappa}{l_\text{BS}})}
\approx (K-1){l_\text{BS}^2}\frac{\sin^2(\pi dK\frac{\kappa}{l_\text{BS}})}{(\pi d\kappa)^2} , \label{E:approx-f}
\end{align}
where the first approximation follows from $\eta_{\text{P}}\approx1$ and the second approximation comes from the first-order Taylor approximation of $\sin(\cdot)$.

Upon substituting \eqref{E:approx-f} and $\eta_{\text{P}}\approx1$  into \eqref{E:approx-ratio}, we obtain
\begin{align} \label{Eq.simpleRatio}
\frac{g_{\text{F},k}}{g_{\text{P},k}}\approx\frac{l_\text{BS}(\eta_\text{F}-1)^2[(2K-3)
\eta_\text{F}^2+2\eta_\text{F}+1]}{[1+(K-1)\eta_\text{F}^2] [K-\frac{l_\text{BS}^2{\sin^2(\pi dK\frac{\kappa}{l_\text{BS}})}}{K(\pi d\kappa)^2}] }.
\end{align}

\begin{figure}
\centering
\includegraphics[width=0.8\textwidth]{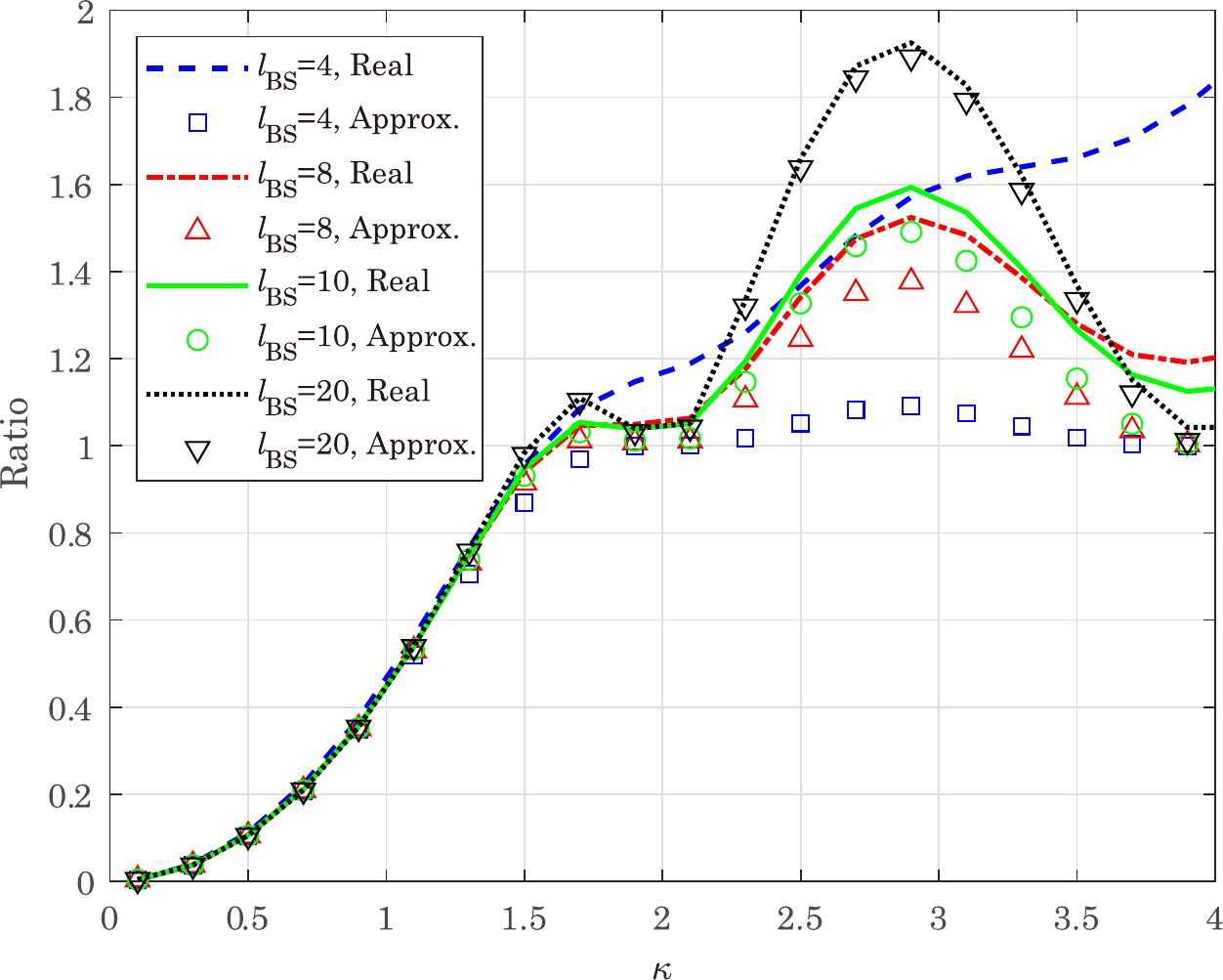}
\caption{$\frac{g_{\text{F},k}}{g_{\text{P},k}}$ vs. $\kappa$ for different setting of $l_{\text{BS}}$, $M_{\text{P}} = 16$, $d=\frac{1}{2}$, and $K=l_{\text{BS}}$.}
\label{Fig:ratio_compare1}
\end{figure}

For the sanity check of the approximation by (\ref{Eq.simpleRatio}), we compare the real value of $\frac{g_{\text{F},k}}{g_{\text{P},k}}$ obtained by simulations and the approximation by (\ref{Eq.simpleRatio}) in Fig.~\ref{Fig:ratio_compare1}, where $M_\text{P}=16$, $d=\frac{1}{2}$, and $K=l_\text{BS}$. We consider different settings of $l_\text{BS}$ and $\kappa$. It is shown that in the interested regime, i.e., the values of $\kappa$ making $\frac{g_{\text{F},k}}{g_{\text{P},k}}$ close to 1, (\ref{Eq.simpleRatio}) approximates well for different settings of $l_\text{BS}$, and the approximation accuracy improves with $l_\text{BS}$, e.g., more than 8.

Based on (\ref{Eq.simpleRatio}), we next examine the condition to determine which HB structure performs better. We consider the following two cases according to whether the system is fully multiplexing or not.

\subsubsection{\underline{Full-multiplexing: $K=l_\text{BS}$}}
In the full-multiplexing mode, the number of the served users equals to the number of RF chains, i.e., $K=l_\text{BS}$. Then, (\ref{Eq.simpleRatio}) reduces to
\begin{align}
\frac{g_{\text{F},k}}{g_{\text{P},k}}\approx\frac{(\eta_\text{F}-1)^2[(2l_\text{BS}-3)
\eta_\text{F}^2+2\eta_\text{F}+1]}{[1+(l_\text{BS}-1)\eta_\text{F}^2](1-\eta_\text{F}^2)},
\label{Eq.ratioAsy}
\end{align}
where $\eta_\text{F} = \frac{\sin(\pi d\kappa)}{\pi d\kappa}$, as defined below \eqref{E:g-full-approx}, is used.

Based on (\ref{Eq.ratioAsy}), to achieve $\frac{g_{\text{F},k}}{g_{\text{P},k}}>1$, we need to satisfy the following condition:
\begin{align}
    l_\text{BS}(1-3\eta_\text{F})>4(1-\eta_\text{F}),\ \eta_\text{F}\neq 0,
\end{align}
from which we obtain the following proposition.

\emph{\textbf{Proposition 1:} Given the analog precoder formed by steering vectors towards UEs, the following operating region needs to be satisfied under the full-multiplexing mode, so that the full-connection structure performs better than the partial-connection structure:}
\begin{align} \label{E:condition1}
    \l_\text{BS}>\frac{4(1-\eta_\text{F})}{1-3\eta_\text{F}},\ \eta_\text{F}<\frac{1}{3},\ \eta_\text{F}\neq 0.
\end{align}

\emph{\textbf{Remark 1:}} Proposition 1 indicates that the full-connection structure is always inferior to the partial-connection structure when $\eta_\text{F}>\frac{1}{3}$, corresponding to the normalized angle separation between UEs $\kappa < 1.45$ for the antenna spacing $d=\frac{1}{2}$ wavelength. For $\kappa > 1.45$, in order to ensure the full-connection structure is better, more RF chains are required. For instance, when $\kappa = 1.6$ and $d=\frac{1}{2}$, i.e., $\eta_\text{F} = 0.23$, $\l_\text{BS}$ should be larger than $10$.

%

To do the sanity check for proposition 1,  we plot the values of $\frac{g_{\text{F},k}}{g_{\text{P},k}}$ for different $\eta_\text{F}$, where both the real value obtained by simulations and the approximation in \eqref{Eq.ratioAsy} are exhibited, $M_\text{P}=16$, $d=\frac{1}{2}$, and $K=l_\text{BS}$. It is shown that $\frac{g_{\text{F},k}}{g_{\text{P},k}}<1$ always holds when $\eta_\text{F} > \frac{1}{3}$. For $\eta_\text{F} < \frac{1}{3}$, $\frac{g_{\text{F},k}}{g_{\text{P},k}}$ may be still less than 1, e.g., when $l_\text{BS}=16$ and $\eta_\text{F}$ is close to $\frac{1}{3}$. Yet, increasing $l_\text{BS}$ can improve $\frac{g_{\text{F},k}}{g_{\text{P},k}}$, for instance, which is always not smaller than $1$ when $l_\text{BS}=128$ for any $\eta_\text{F} < \frac{1}{3}$. The results are consistent with Proposition~1.
In addition, one can find that $\frac{g_{\text{F},k}}{g_{\text{P},k}} = 1$ holds when $\eta_\text{F}=0$ for all $l_\text{BS}$. This can be readily verified from \eqref{Eq.ratioAsy}.


\begin{figure}
\centering
\includegraphics[width=0.8\textwidth]{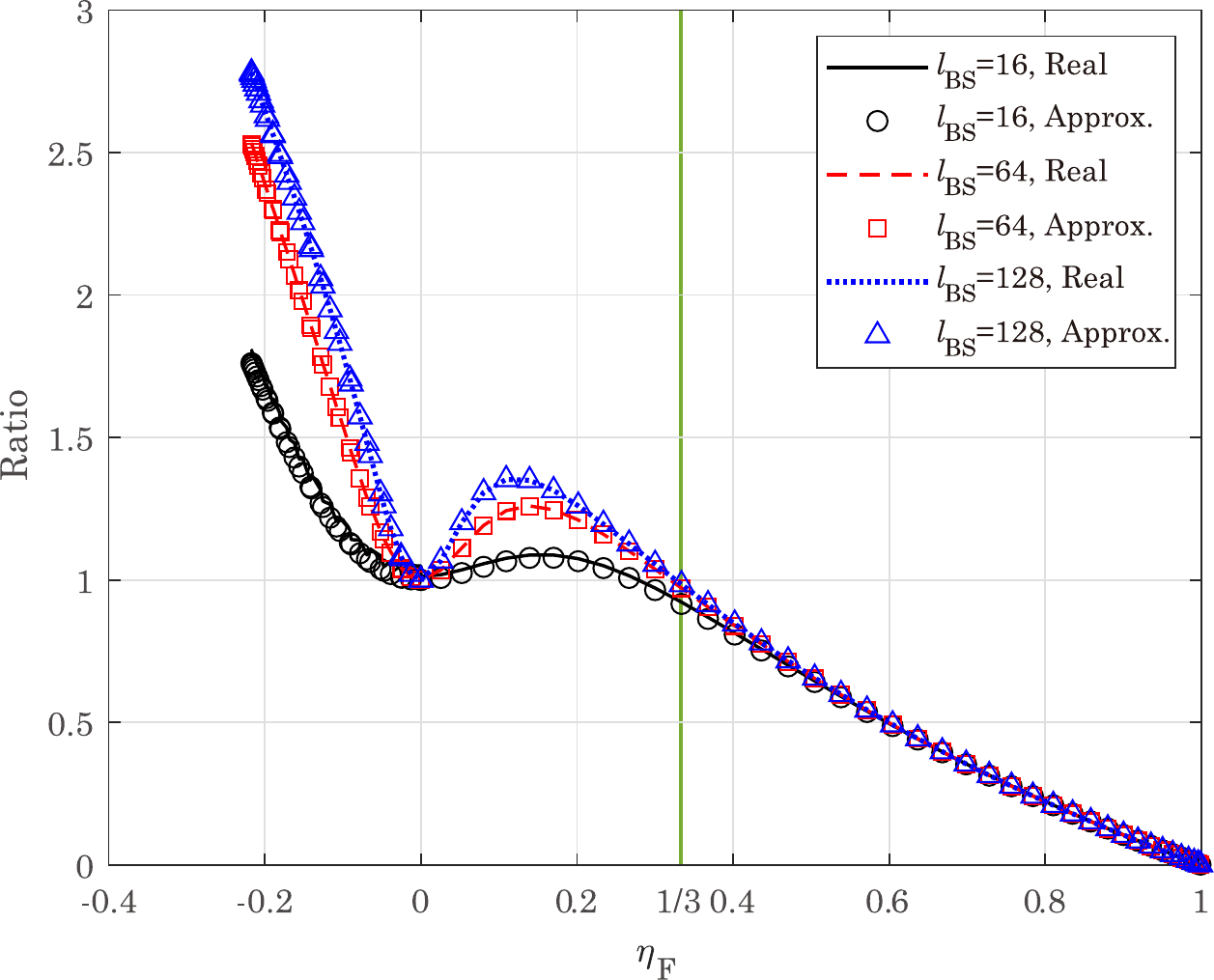}
\caption{$\frac{g_{\text{F},k}}{g_{\text{P},k}}$ vs. $\eta_\text{F}$ for different $l_{\text{BS}}$, $M_{\text{P}} = 16$, and $K=l_{\text{BS}}$.}
\label{Fig:ratioEta_asymptotic_Msub=4}
\end{figure}



\subsubsection{\underline{Arbitrary $K\leq l_\text{BS}$}}
We now consider a general case with the number of scheduled users not larger than the number of RF chains, i.e., $K\leq l_\text{BS}$. For notational simplicity, define $\rho\triangleq \frac{K}{l_\text{BS}}$, where $\frac{1}{l_\text{BS}}\leq \rho\leq 1$. Substitute $\rho$ into (\ref{Eq.simpleRatio}), we acquire that
\begin{align} \label{E:g-K}
&\frac{g_{\text{F},k}}{g_{\text{P},k}}\approx\frac{(\eta_\text{F}-1)^2[(2\rho l_\text{BS}-3)\eta_\text{F}^2+2\eta_\text{F}+1]}{\rho[1+(\rho l_\text{BS}-1)\eta_\text{F}^2](1-\frac{\sin^2(\pi d\rho \kappa)}{(\pi d\rho\kappa)^2})}. 
\end{align}

We first examine the value of $\frac{g_{\text{F},k}}{g_{\text{P},k}}$ when $\eta_{\text{F}} = 0$, which requires $\sin(\pi d \kappa)=0$ and $\kappa \neq 0$ according to the definition of $\eta_{\text{F}}$ as defined below \eqref{E:g-full-approx}, i.e., $\kappa = \frac{n}{d}$ should hold for $n\in\mathbb{Z}^+$. In this case, \eqref{E:g-K} reduces to
\begin{align}
\frac{g_{\text{F},k}}{g_{\text{P},k}}\approx\frac{1}{\rho(1-\frac{\sin^2(\pi n \rho)}{(\pi n \rho)^2})}\geq \frac{1}{\rho}\geq 1.
\end{align}

When $\eta_{\text{F}} \neq 0$, i.e., $\kappa \neq \frac{n}{d}, \forall n\in\mathbb{Z}^+$, the condition to ensure $\frac{g_{\text{F},k}}{g_{\text{P},k}}\geq 1$ can be derived from \eqref{E:g-K}~as
\begin{align}
    (2-A)l_\text{BS}\geq \frac{(3-A)\eta_\text{F}^2-2\eta_\text{F}-1+A}{\eta_\text{F}^2\rho},
\label{Eq.boundOriginal}
\end{align}
where $A\triangleq \frac{1-\frac{\sin^2(\pi d\rho \kappa)}{(\pi d\rho \kappa)^2}}{(1-\eta_\text{F})^2}\rho$.
In Appendix~\ref{A:proof1}, we prove that the condition \eqref{Eq.boundOriginal} is infeasible when $A > 2$. When $A < 2$, \eqref{Eq.boundOriginal} can be transformed into the condition for $l_\text{BS}$ as
\begin{align}
    l_\text{BS}\geq \frac{(3-A)\eta_\text{F}^2-2\eta_\text{F}-1+A}{\eta_\text{F}^2\rho(2-A)},\ A<2.
\label{Eq.feasible}
\end{align}
When $A=2$, (\ref{Eq.boundOriginal}) becomes $0\geq \frac{(\eta_\text{F}-1)^2}{\eta_\text{F}^2\rho}$, which is infeasible except for $\eta_{\text{F}}=1$ and thus indicates that the full-connection structure cannot perform better in this case.

The above analysis leads to the following proposition.

\emph{\textbf{Proposition 2:} Given analog precoder formed by steering vectors towards UEs, the following two operating regions need to be satisfied when $K\leq l_\text{BS}$, so that the full-connection structure performs better than the partial-connection structure:}
\begin{align} \label{E:condition2}
    \begin{cases}
          l_\text{BS}\geq \frac{(3-A)\eta_\text{F}^2-2\eta_\text{F}-1+A}{\eta_\text{F}^2\rho(2-A)} \triangleq l_\text{BS}^{\text{th}}, &\ A<2, \kappa \neq \frac{n}{d}, \forall n\in\mathbb{Z}^+, \\
          \text{any}\ l_\text{BS}, & \ \kappa = \frac{n}{d}, \forall n\in\mathbb{Z}^+.
    \end{cases}
\end{align}

\emph{\textbf{Remark 2:}} When $\rho=1$, i.e., $K=l_\text{BS}$, we have $A = \frac{1+\eta_{\text{F}}}{1-\eta_{\text{F}}}$, with which condition \eqref{E:condition2} reduces to condition \eqref{E:condition1} in Proposition~1.

\begin{figure}
\centering
\includegraphics[width=0.8\textwidth]{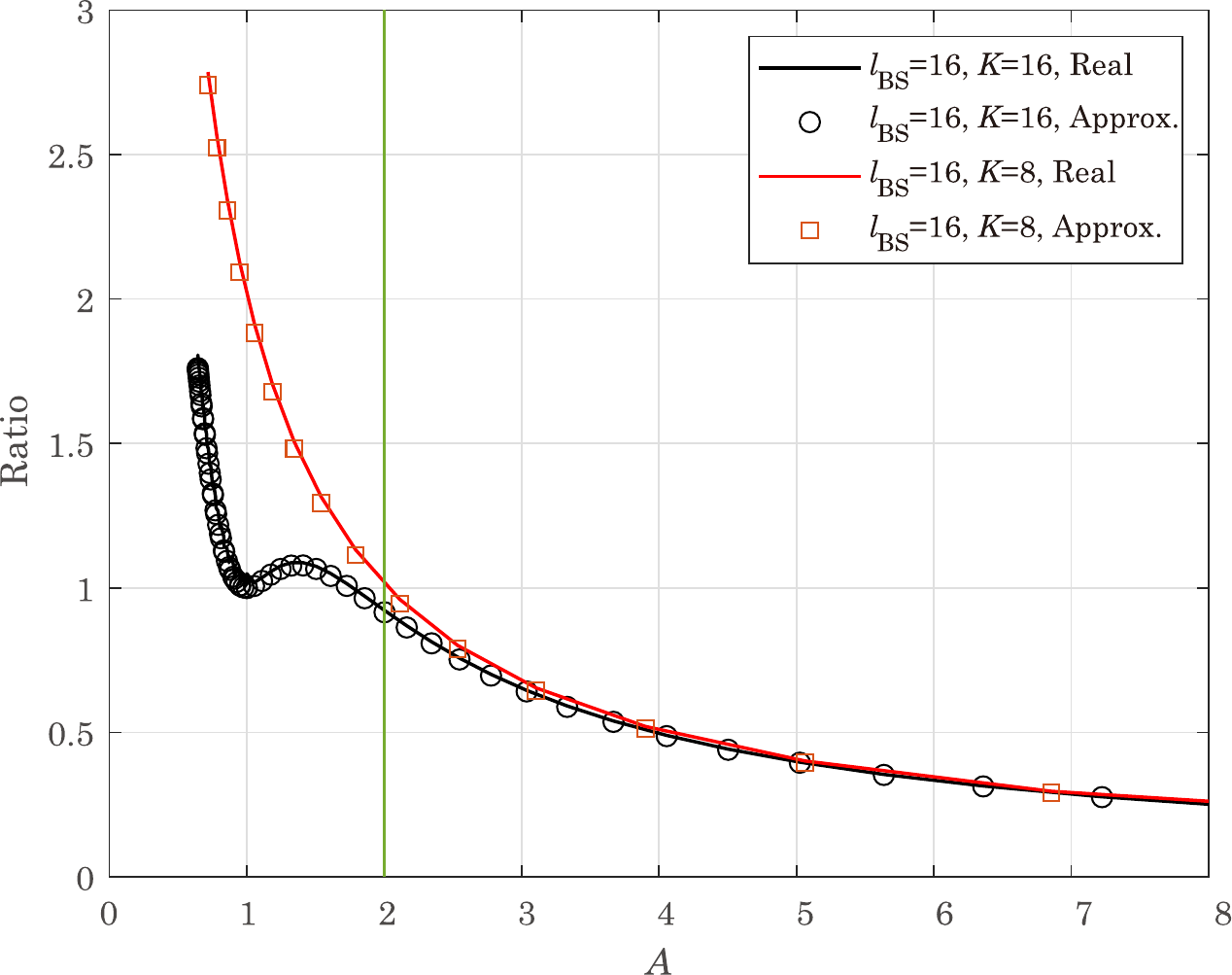}
\caption{$\frac{g_{\text{F},k}}{g_{\text{P},k}}$ vs. $A$, $M_{\text{P}} = 16$, $l_{\text{BS}} = 16$, and $K=8$ and $16$ (i.e., $\rho = 0.5$ and $1$).}
\label{Fig:ratioEta_non}
\end{figure}

In Fig.~\ref{Fig:ratioEta_non}, we plot the values of $\frac{g_{\text{F},k}}{g_{\text{P},k}}$ for different $A$, where both the real value obtained by simulations and the approximation in \eqref{E:g-K} are exhibited, $M_\text{P}=16$, $d=\frac{1}{2}$, and $l_\text{BS} = 16$. We consider $K = 8$ and $16$, corresponding to $\rho = 0.5$ and $1$, respectively. When $\rho = 1$, we can find that $\frac{g_{\text{F},k}}{g_{\text{P},k}} < 1$ for $A \geq 2$, while $\frac{g_{\text{F},k}}{g_{\text{P},k}}$ is not always larger than 1 when $A < 2$, depending on whether the condition $l_\text{BS}\geq l_\text{BS}^{\text{th}}$ is satisfied or not. This coincides with Proposition 2. When $\rho = 0.5$, we can find that $\frac{g_{\text{F},k}}{g_{\text{P},k}} > 1 $ always holds when $A < 2$, which is consistent with the analysis in Remark 2 for small $\rho$.


%
%
%

\section{Conclusions}
This article compared the downlink performance of two hybrid beamforming structures, namely full- and partial-connections, where it is assumed that the angular spread for the mmWave channels is small so that the analog precoder is formed towards users. Given the analog precoder and zero-forcing digital precoder, we developped an upper bound for the analog-and-digital precoded channel gain of users, based on which the precoded channel gain ratio of the full-connection structure over the partial-connection structure was analyzed. We find that the full-connection structure does not always achieve a larger precoded channel gain, which depends on both the angular separation between users and the number of RF chains. We revealed the regimes suitable for the two structures, which are validated by simulations.

\appendices
\section{Proof of Lemma 1} \label{A:proof-lemma1}
Recall that $h_i =\|\tilde{\mathbf{U}}_i^\dag\bar{\mathbf{h}}_i\|^2$ and $\tilde{\mathbf{U}}_i$ is semi-unitary that are orthogonal to the subspace of $\tilde{\mathbf{H}}_i$. Let ${\mathbf{U}}_i$ denote a semi-unitary matrix that spans the subspace of $\tilde{\mathbf{H}}_i$, which can be constructed in the form ${\mathbf{U}}_i=\left[\frac{\bar{\mathbf{h}}_j}{\|\bar{\mathbf{h}}_j\|},{\mathbf{U}}_{i,j}\right]$, $j\neq i$, where ${\mathbf{U}}_{i,j}$ is orthogonal to $\frac{\bar{\mathbf{h}}_j}{\|\bar{\mathbf{h}}_j\|}$.
Then, we can obtain that
\begin{align}
  h_i &=\|\tilde{\mathbf{U}}_i^\dag\bar{\mathbf{h}}_i\|^2 = \bar{\mathbf{h}}_i^\dag\tilde{\mathbf{U}}_i\tilde{\mathbf{U}}_i^\dag\bar{\mathbf{h}}_i = \bar{\mathbf{h}}_i^\dag\left(\mathbf{I}-{\mathbf{U}}_i{\mathbf{U}}_i^\dag\right)\bar{\mathbf{h}}_i \nonumber\\
  &=  \|\bar{\mathbf{h}}_i\|^2 - \bar{\mathbf{h}}_i^\dag \left[\frac{\bar{\mathbf{h}}_j}{\|\bar{\mathbf{h}}_j\|},{\mathbf{U}}_{i,j}\right]
  \left[\frac{\bar{\mathbf{h}}_j}{\|\bar{\mathbf{h}}_j\|},{\mathbf{U}}_{i,j}\right]^\dag \bar{\mathbf{h}}_i\nonumber\\
  &\leq \|\bar{\mathbf{h}}_i\|^2 - \frac{|\bar{\mathbf{h}}_j^\dag\bar{\mathbf{h}}_i|^2}{\|\bar{\mathbf{h}}_j\|^2}.
\end{align}

\section{Infeasibility of \eqref{Eq.boundOriginal} for $A > 2$} \label{A:proof1}

When $A>2$, the left-hand side of \eqref{Eq.boundOriginal} is negative. We next prove that the right-hand side of \eqref{Eq.boundOriginal} is non-negative for $A>2$, making \eqref{Eq.boundOriginal} infeasible.

First, when $2<A<3$, the numerator of the right-hand side of \eqref{Eq.boundOriginal}, denoted by $\Omega(\eta_\text{F})=(3-A)\eta_\text{F}^2-2\eta_\text{F}-1+A$, is monotonically decreasing with $\eta_\text{F}$, because its first-order derivation is $2(3-A)\eta_{\text{F}}-2\leq0$ by recalling that $\eta_\text{F}\in[-0.22,1]$. Then, $\Omega(\eta_\text{F})$ satisfies
\begin{align}
\Omega(\eta_\text{F})\geq \Omega(1)=0.
\end{align}

Second, when $A=3$, we have $\Omega(\eta_\text{F})=-2\eta_\text{F}+2\geq0$.

Third, when $A>3$, in $\Omega(\eta_\text{F})$ the term $(3-A)\eta_\text{F}^2$ is lower bounded by $3-A$, the term $-2\eta_\text{F}$ is lower bounded by $-2$, and thus $\Omega(\eta_\text{F})$ is lower bounded by $\Omega(\eta_\text{F})\geq (3-A)-2-1+A=0$.

In summary, $\Omega(\eta_\text{F})$ is non-negative for $A>2$, which completes the proof.

\bibliography{IEEEabrv,MIMO}

\end{document}